# Ultra-high performance microwave spectral filters using optical microcombs

**David J. Moss**

Optical Sciences Centre, Swinburne University of Technology, Hawthorn, VIC 3122, Australia

**dmoss@swin.edu.au**

**Keywords**: Microwave photonic filters, optical microcombs, integrated optics.

**Abstract**

Microwave photonic (MWP) filters are essential components in microwave systems due to their wide bandwidth, low loss, and immunity to electromagnetic interference. A sharp transition band is critical for precise spectral shaping and interference suppression, yet conventional MWP filters face challenges in achieving both sharp transitions and high reconfigurability. Adaptive MWP filters with sharp transition based on a transversal filter structure using an optical microcomb source are demonstrated in this paper. Four different types of single-band MWP filters with roll-off rates up to ~32.6 dB/GHz and a minimum shape factor of ~1.15 are achieved. In addition, simply through designing tap coefficients, band-pass filters with tunable centre frequencies ranging from 5 GHz to 15 GHz and dual-band MWP filters with various filter response are demonstrated without changing any hardware, where sharp transition is also validated. The adaptive filters with sharp transition presented in this paper offer a stable and highly reconfigurable solution for applications requiring stringent spectral selectivity, such as next-generation wireless networks, high-resolution radar imaging, and advanced biomedical photonic sensing.

## 1. Introduction

Microwave photonic (MWP) filters have emerged as a crucial technology in modern communication systems, radar, and high-speed signal processing due to their ability to process signals with wide bandwidth, low loss, and immunity to electromagnetic interference [1, 2]. These advantages make MWP filters a superior alternative to traditional electronic filters. It is critical to achieve MWP filters with sharp transitions, since it facilitates precise spectral shaping and enhanced suppression of out-of-band interference, thereby improving the accuracy of desired signal extraction [3].

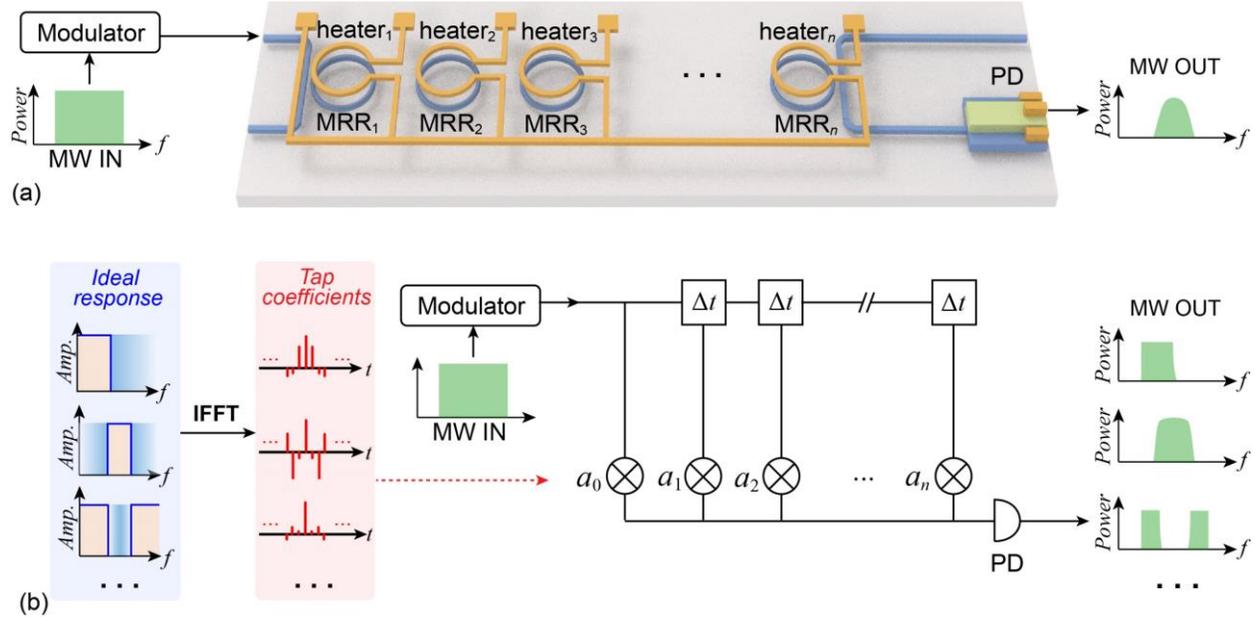

**Figure 1**. Schematic of microwave photonic (MWP) filters with high roll-off performance. (a) Schematic of an infinite impulse response (IIR) MWP filter based on cascaded microring resonators (MRRs). MW: microwave. PD: photodetector. (b) Schematic of an FIR filter based on the transversal filter structure with high reconfigurability. IFFT: inverse fast Fourier transform.

Conventional MWP filters often face inherent challenges in simultaneously achieving high roll-off performance, high reconfigurability, and broad tuning bandwidth. MWP filters can be categorized into infinite impulse response (IIR) MWP filters and finite impulse response (FIR) MWP filters [4]. IIR MWP filters are typically implemented using optical resonators [5-9], optical delay lines [10], optical-electrical feedback loops [11], ring-assisted Mach-Zehnder Interferometers (MZIs) [12, 13], or stimulated Brillouin scattering in waveguides [14, 15], which allow for the realization of high-quality factor filtering responses, enabling sharp roll-off characteristics. While IIR MWP filters exhibit superior spectral selectivity, they are with low reconfigurability and suffer from instability as well as sensitivity to fabrication imperfections [6]. One of the approaches to achieving high roll-off performance in MWP filters is the use of cascaded microresonators (**Figure 1(a)**). Multiple coupled resonators with high quality factors are cascaded to achieve sharp filtering responses, which introduces significant design and fabrication challenges [9]. The resonance characteristics of each microresonator must be precisely controlled, and thermal variations as well as fabrication-induced resonance mismatches can easily degrade filter performance.

FIR MWP filters, typically implemented via transversal filter structures (**Figure 1(b)**) [16-19], offer high flexibility and stability due to their inherent non-recursive nature. They use a finite number of discrete taps to perform weighted summation, allowing easy reconfiguration. However, a limitation of FIR MWP filters is their relatively slow roll-off rate, as achieving sharp spectral transitions requires a large number of taps, which can be difficult to implement in traditional configurations based on discrete laser arrays [32] and fibre Bragg gratings (FBGs) [3].

Recently, optical microcombs [20-22], which are optical frequency combs generated by micro-scale resonators, offer distinct advantages for FIR MWP filters, as they can provide a larger number of discrete wavelengths while also having a compact device footprint [23-25]. This makes them an ideal candidate for realizing MWP filters with enhanced roll-off characteristics. Each comb line serves as an independent tap in the transversal filter structure, and the amplitude of each tap can be designed to enable filter responses with steep roll-off performance. Unlike IIR MWP filters, which rely on feedback loops and resonators, the microcomb-based approach does not suffer from inherent stability or sensitivity to fabrication imperfections. Another key advantage of the microcomb-based MWP filters is the high reconfigurability. The filter shape can be modified simply by designing different tap coefficients to adapt to different filtering requirements without changing any hardware.

In this paper, we demonstrate ultrahigh-performance adaptive MWP filters based on an optical microcomb source. We achieve four single-band filters with different filter shapes, which have extremely sharp roll-off performance with roll-off rates up to ~32.6 GHz and a minimum 20-dB shape factor of ~1.15. To describe the tunability of the filters, band-pass filters with tunable centre frequencies ranging from 5 GHz – 15 GHz are demonstrated. In addition, we present dual-band filters with various filter response by using the same configuration to verify the reconfigurability, where high roll-off performance is also validated. The ultrahigh-performance adaptive filters presented in this paper open new avenues for the application of MWP filters in systems requiring high spectral selectivity, such as wireless communication, radar, and biomedical imaging systems, where sharp roll-off performance and different filter response are necessary.

## 2. Principle

**Figure 2(a)** shows the schematic diagram of an ultrahigh-performance MWP filter based on a soliton crystal microcomb. The soliton crystal microcomb is generated by an integrated microring resonator (MRR) and the schematic of microcomb generation is shown in **Figure 2(b)**. The MRR is pumped by a continuous-wave (CW) laser, which is amplified by an erbium-doped fibre amplifier (EDFA). A polarization controller is used to adjust the polarization, optimizing the power coupled to the MRR. When the pump laser is precisely tuned, it transitions from a blue-detuned to a red-detuned regime, which leads to the generation of the soliton crystal microcomb with a free spectral range (FSR) of ~49 GHz. The generated microcomb, which serves as a multi-wavelength optical source, is fed into an intensity modulator (IM), yielding replicas of the input microwave signal. Then the modulated optical signal goes through a spool of single-mode fibre (SMF), where a time delay between adjacent wavelength channels $\Delta t$ in **Figure 1** is introduced. Next, the delayed signal is shaped by a Waveshaper to achieve the desired tap coefficients. The calculated tap coefficients of low-pass, band-pass, high-pass, and band-stop filters are shown in **Figure 2(c)**. Instead of employing equal tap coefficients to realize low-pass filters and with additional modifications to shape the filter response in our previous work [18, 26], we directly design the tap coefficients by applying inverse fast Fourier transform (IFFT) to the ideal filters' radio-frequency (RF) amplitude response. This approach enables improved roll-off performance by ensuring that the obtained filter shapes more accurately align with the ideal response. Finally, the delayed and weighted taps are summed upon photodetection and converted to microwave signals at the output. A vector network analyzer (VNA) is used to measure the RF response of the system. **Figure 2(d)** shows the simulated RF amplitude response of low-pass, band-pass, high-pass, and band-stop filters.

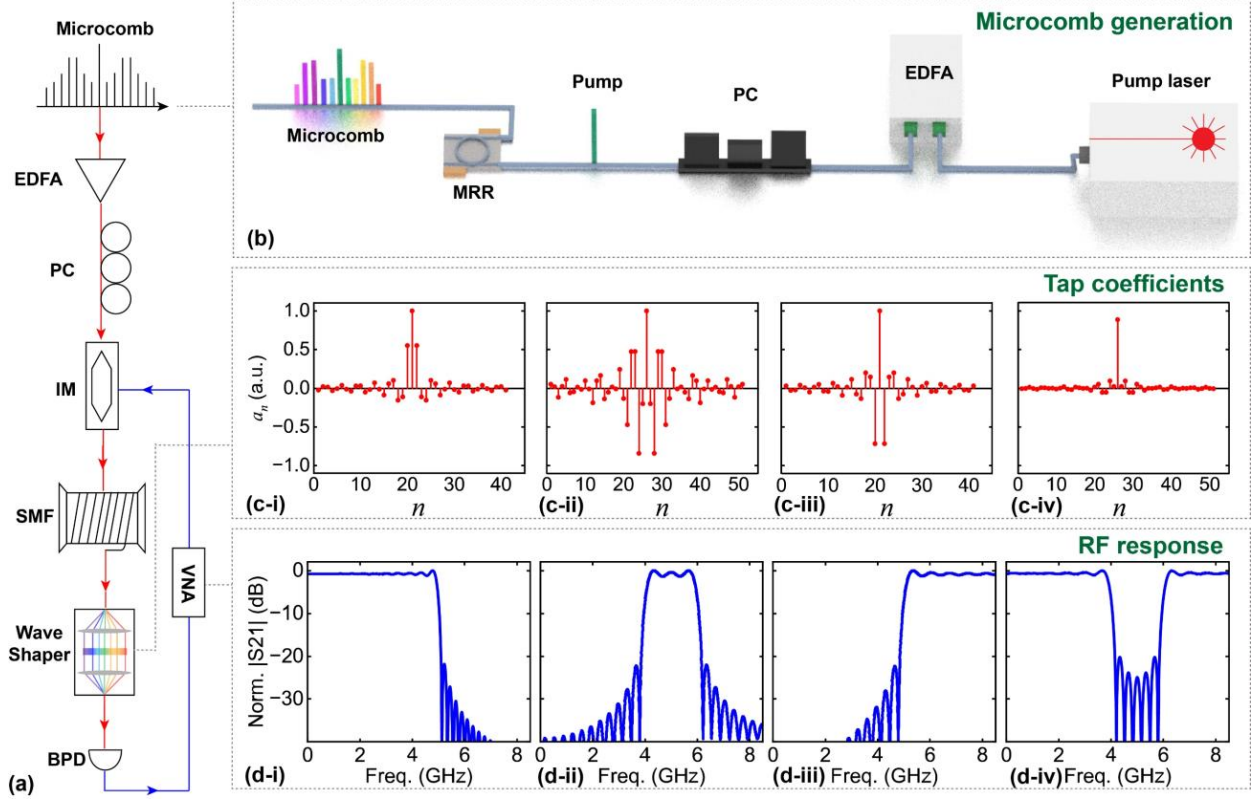

**Figure 2**. Schematic configuration of ultrahigh-performance adaptive MWP filters. (a) Schematic of an ultrahigh-performance microwave photonic (MWP) filter based on a soliton crystal microcomb. EDFA: erbium-doped fibre amplifier. PC: polarization controller. IM: Intensity modulator. SMF: single-mode fibre. BPD: balanced photo detector. VNA: vector network analyzer. (b) Schematic of soliton crystal microcomb generation. MRR: microring resonator. (c) Calculated tap coefficients of (i) low-pass filter, (ii) band-pass filter, (iii) high-pass filter, and (iv) band-stop filter. (d) Simulated RF amplitude response of (i) low-pass filter, (ii) band-pass filter, (iii) high-pass filter, and (iv) band-stop filter.

The transfer function of the MWP transversal filter system in **Figure 2(a)** can be described as [27]

$$H(\omega) = \sum_{n=0}^{M-1} a_n e^{-j\omega n \Delta t},$$ (1)

where $M$ is the number of taps, $a_n$ ($n = 0, 1, 2, \ldots, M-1$) is the tap coefficient of the $n^{\text{th}}$ tap, and $\omega$ is the angular frequency of the input microwave signal. By designing $a_n$ ($n = 0, 1, 2, \ldots, M-1$), different filter shapes can be achieved without changing the hardware. The time delay $\Delta t$ is determined by the FSR of microcomb $\Delta\lambda$ and the dispersion of the SMF, which can be further expressed as [20]

$$\Delta t = \Delta\lambda \times L \times D_2, \tag{2}$$

where $L$ is the fibre length and $D_2$ is the second-order dispersion (SOD) parameter of the dispersive module. As MWP transversal filters have a finite impulse response, they exhibit a periodic spectral response, and the FSR can be calculated by $FSR_{MW} = 1/\Delta t$. The operation bandwidth (*i.e.*, the Nyquist frequency) is half of $FSR_{MW}$, which can be expressed as [20]

$$OB = \frac{1}{2\,\Delta t}. \tag{3}$$

The roll-off rate and shape factor are two important metrics to demonstrate the roll-off performance of MWP filters. The roll-off rate (*ROR*) quantifies how rapidly the filters transition from passbands to stopbands, which can be described as [19]

$$ROR = \frac{A}{|f_{-3\mathrm{dB}} - f_{ref}|}, \tag{4}$$

where A is the attenuation from the -3 dB point in the transition band to a reference point, and $f_{3dB}$ and $f_{ref}$ are the frequencies at $-3$ dB point and the reference point, respectively. A higher roll-off rate indicates better stop-band suppression and less interference. The shape factor (*SF*) describes filters' spectral selectivity and is defined as the ratio of bandwidth at a reference level to the bandwidth at the $-3$dB level. The shape factor is given by [7]

$$SF = BW_{ref} / BW_{-3\mathrm{dB}}, \tag{5}$$

where $BW_{ref}$ and $BW_{\text{-3dB}}$ are the bandwidth at a reference level and -3 dB level, respectively. A smaller shape factor (closer to 1) indicates a sharper transition and better spectral selectivity.

## 3. Experimental demonstration

### 3.1 Soliton crystal microcomb

Soliton crystal microcomb represents a unique and robust class of optical frequency comb, operating in a highly coherent state that is enabled by optical parametric oscillation within an integrated MRR. The MRR used here is fabricated on a complementary metal-oxide-semiconductor (CMOS)-compatible doped silica glass platform [28, 29], with a quality (Q) factor of approximately 1.9 million (**Section S1**, Supporting Information) and a radius of ~592 μm as shown in **Figure 3(a)**, yielding a FSR of ~49 GHz (or ~0.4 nm). This narrow FSR results in a large number of available wavelengths in the telecommunications C-band.

The soliton crystal microcomb is characterized by deterministic formation, facilitated by mode-crossing-induced background waves and the Kerr nonlinearity, combined with high intracavity power [30]. The soliton crystal microcomb is generated by using avoided mode crossing (AMX), which is shown in **Figure 3(b)** and obtained by employing the setup described in **Section S1** (Supporting Information) [31]. Unlike dissipative Kerr solitons (DKS), soliton crystal microcomb exhibits minimal intracavity energy variation during its generation, thereby eliminating the need for complex tuning techniques. The initiation of the soliton crystal microcomb is achieved via a simple pump wavelength sweeping method [32], where a CW pump laser (Yenista Optics), amplified to 32.1 dBm by an EDFA (IdealPhotonics), is manually swept in wavelength from blue to red until modulation instability oscillations emerge. With the detuning between the pump and the MRR's resonance slipping farther, a stable soliton crystal oscillation state is achieved, generating over 90 wavelength channels across the C band at a pump wavelength of 1551.3 nm, as shown in **Figure 3(c)**. As can be seen, the soliton crystal microcomb exhibits scalloped-shaped spectra that, although it has been regarded as a potential limitation for practical applications, has in fact been shown not to present a significant drawback. This is partly attributed to the higher power conversion efficiency of soliton crystal microcombs compared with single soliton states [20]. In addition, the soliton crystal microcomb demonstrated remarkable long-term power stability, with a measured relative standard deviation of -14 dB over 66 hours [33], confirming its suitability for advanced photonic signal processing applications.

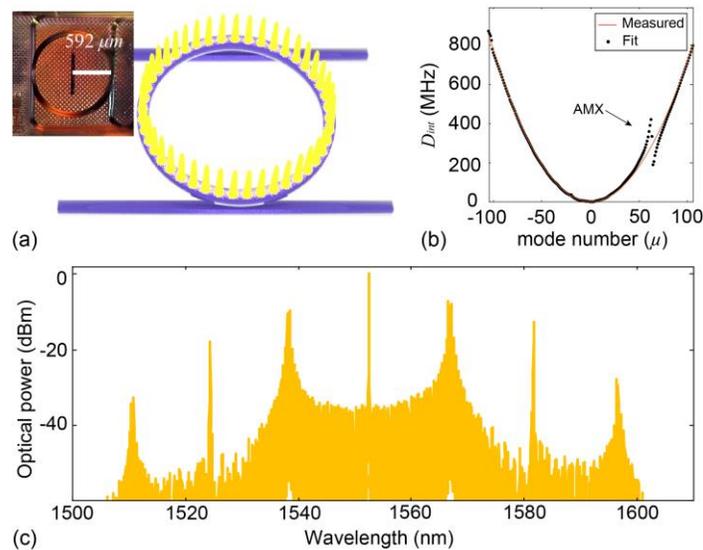

**Figure 3**. Soliton crystal microcomb. (a) Schematic representation and the image (upper inset) of the integrated MRR. (b) Measured integrated dispersion $D_{int}$ of the integrated MRR showing an avoided mode crossing (AMX) location of ~1580.2 nm. (c) Optical spectrum of the generated soliton crystal microcomb with a free spectral range (FSR) of ~49 GHz.

### 3.2 Ultrahigh-performance adaptive MWP filters

The MWP filters with sharp roll-off performance and diverse filter shapes were experimentally demonstrated. The input microwave signal was modulated onto the generated microcomb (**Figure 3(c)**) yielding microwave replicas. Then the modulated optical signal went through a spool of SMF with a length $L$ of ~8.2 km to provide the time delay $\Delta t$. The SOD parameter $D_2$ of the SMF was ~17.4 ps $nm^{-1}$ $km^{-1}$, which corresponded to a $\Delta t$ of ~56.5 ps and an operation bandwidth $OB$ of ~8.84 GHz according to **Equation (3)**. To achieve desired tap coefficients $a_n$ ($n = 0, 1, 2, …, M − 1$), the signal was shaped in power via a Waveshaper (Finisar). The Waveshaper also separated the wavelength channels into two categories according to the sign of tap coefficients. The signals were then directed to the two ports of a balanced photodetector (Finisar), where the weighted and delayed taps were summed upon photodetection, and positive as well as negative taps were enabled.

To demonstrate the sharp roll-off performance of the MWP filters, we implemented low-pass and high-pass filters with cut-off frequencies $f_{ct}$ of 5 GHz, as well as band-pass and band-stop filters with centre frequencies $f_c$ of 5 GHz and a pass/stop band of 2 GHz. The calculated tap coefficients and simulated RF amplitude response are shown in **Figures 2(c)** and **(d)**, respectively.

**Figure 4(a)** shows the designed tap coefficients and the measured optical spectra of the shaped microcomb, measured using an optical spectrum analyzer (OSA, Anritsu), for the four different types of MWP filters. Further details on the tap number for proposed MWP filters are discussed in **Section S2** (Supporting Information). A close agreement between the designed tap coefficients (green circles for positive tap coefficients and yellow circles for negative tap coefficients) and the power of the measured microcomb lines (green solid lines for positive tap coefficients and yellow solid lines for negative tap coefficients) was obtained, indicating that the microcomb lines were accurately shaped.

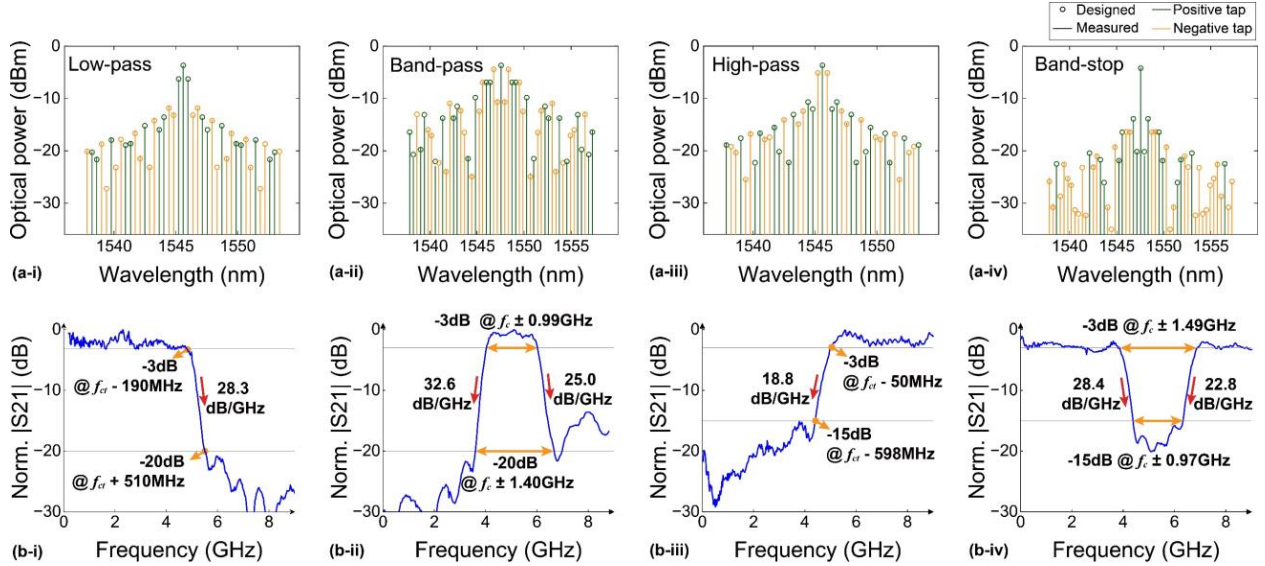

**Figure 4**. Experimental demonstration of MWP filters. (a) Designed tap coefficients (circles) and optical spectra of shaped microcomb (solid lines) for (i) low-pass filter, (ii) band-pass filter, (iii) high-pass filter, and (iv) band-stop filter. Green circles and solid lines for positive tap coefficients and yellow circles and solid lines for negative tap coefficients. (b) Measured RF amplitude response of (i) a low-pass filter with a roll-off rate of ~28.3 GHz and a 20-dB shape factor of ~1.15, (ii) a band-pass filter with a roll-off rate of ~32.6 GHz for low-frequency edge as well as ~25.0 GHz for high-frequency edge and a 20-dB shape factor of ~1.55, (iii) a high-pass filter with a roll-off rate of ~18.8 GHz and a 15-dB shape factor of ~1.17, and (iv) a band-stop filter with a roll-off rate of ~28.4 GHz for low-frequency edge as well as ~22.8 GHz for high-frequency edge and a 15-dB shape factor of ~1.53.

We measured the RF amplitude response of the filters by a VNA (Anritsu) as shown in **Figure 4(b)**. The low-pass filter exhibited a high roll-off rate of ~28.3 dB/GHz with a reference point $f_{ref}$ in **Equation (4)** at −20 dB. The roll-off rates of the band-pass filter [$ROR_l$, $ROR_h$], where $ROR_l$ and $ROR_h$ are defined as the roll-off rates for low- and high-frequency edges, were ~[32.6, 25.0] dB/GHz at $f_{ref}$ = −20 dB. For the high-pass filter, the roll-off rate was ~18.8 dB/GHz at $f_{ref}$ = −15 dB. The band-stop filter exhibited roll-off rates of ~[28.4, 22.8] dB/GHz at $f_{ref}$ = −15 dB. The $BW$-3dB and $BW$-20dB of the low-pass filter (**Figure 4(b-i)**) were $f_{ct}$ − 190 MHz and $f_{ct}$ + 510 MHz, respectively, corresponding to a 20 dB shape factor of ~1.15. The shape factors of the band-pass, high-pass, and band-stop filters (**Fig. 4(b-ii)–(b-iv)**) were ~1.55, ~1.17, and ~1.53, respectively.

The results demonstrate that the proposed filters have achieved a sharp roll-off performance. In our previous work [26, 34], low-pass filters were initially realized with equal tap coefficients, followed by refinement of the tap coefficients to tailor the filter response. Instead of employing this method, we directly obtained the tap coefficients by applying IFFT to the ideal RF amplitude response here. Compared to our previous work [26, 34], this method ensured that the system response more closely aligned with the ideal response, which significantly improved the roll-off rate from 5.2 dB/GHz to 32.6 dB/GHz, achieving a shape factor of ~1.15. The sharp roll-off performance is essential for improving signal selectivity and mitigating spectral leakage from adjacent channels in microwave signal processing and data transmission systems. Since the soliton crystal microcomb provided enough number of taps (**Section S2**, Supporting Information), the roll-off performance of the proposed MWP filters was mainly limited by the experimentally induced processing errors, including noise of the soliton crystal microcomb, chirp and uneven response of the IM, high-order dispersion in SMF, shaping errors of the Waveshaper, and noise of the BPD [35]. In our experiments, we employed a two-stage feedback control method (see **Section S3**, Supporting Information) to calibrate the tap coefficients [36], which mitigated the errors resulting from static and slowly varying error sources [37]. Residual errors remained in the system that could not be compensated for via the implemented two-stage feedback control method. These were mainly fast varying errors, such as the amplitude and phase distortions induced by the microcomb and BPD.

To demonstrate the tunability of the MWP filters, the tap coefficients were designed to achieve band-pass filters with different centre frequencies $f_c$. Here we employed a spool of SMF with a fibre length of ~5.1 km, corresponding to an operation bandwidth $OB$ of ~28.7 GHz. We note that the maximum *OB* is limited by half of the FSR of the soliton crystal microcomb. When the operation is beyond half of the FSR (~24.5 GHz), strong crosstalk between adjacent taps occurs.

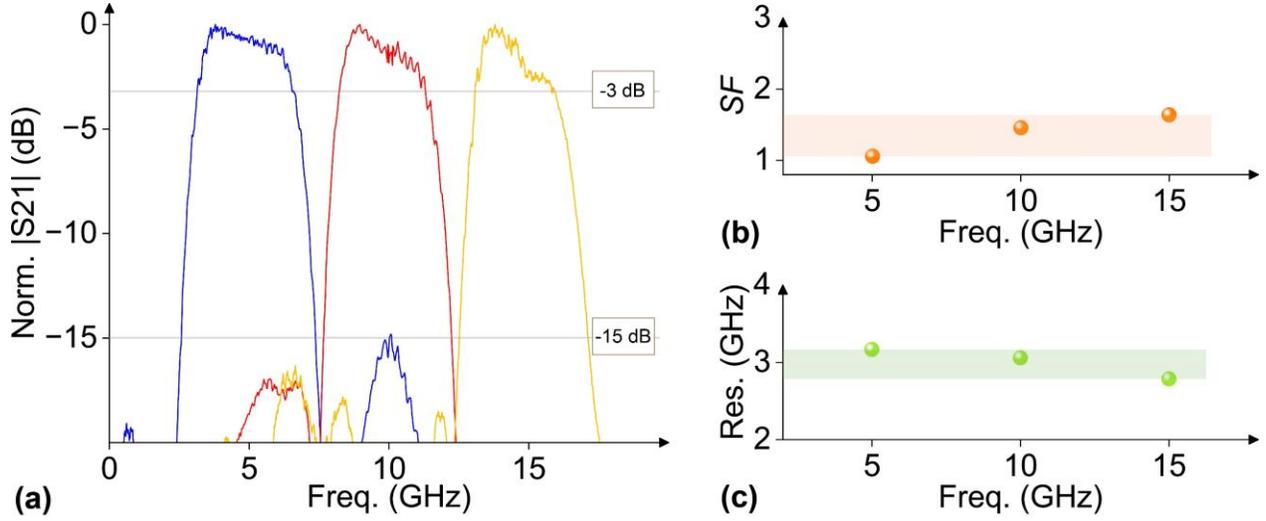

**Figure 5**. Tunability of MWP band-pass filters. (a) Measured RF amplitude response of tunable band-pass filters with centre frequencies of 5 GHz, 10 GHz, and 15 GHz. (b) Features of the tunable band-pass filters' shape factors. (c) Features of the tunable band-pass filters' resolution.

As shown in **Figure 5(a)**, three band-pass filters have been realized, each with a 3-GHz passband bandwidth and tunable centre frequencies ranging from 5 GHz to 15 GHz. The roll-off rates were ~[17.2, 16.1] dB/GHz, ~[21.4, 14.3] dB/GHz, and ~[21.0, 12.9] dB/GHz for the three band-pass filters with centre frequencies at 5 GHz, 10 GHz, and 15 GHz, respectively. With the tuning of the centre frequencies, the filters showed consistent low shape factors ranging from ~1.06 to ~1.64, as can be seen from **Figure 5(b)**. This consistency suggests that the proposed MWP filters keep a sharp transition band without being affected by the centre frequency, which is highly beneficial for applications requiring uniform filter performance across a wide bandwidth. The resolution (3-dB bandwidth) of the filters also remained consistent (**Figure 5(c)**) with a deviation under 210 MHz. The combination of sharp roll-off and tunable centre frequency while maintaining stable resolution reflects that the proposed MWP filter system is well-calibrated and precisely controlled.

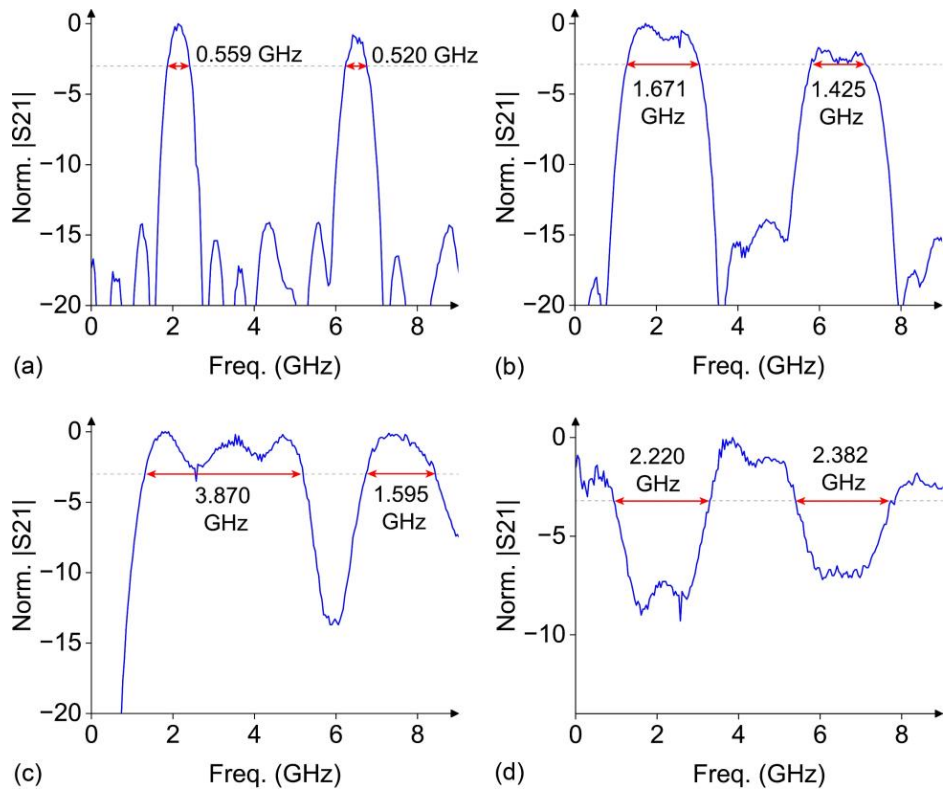

**Figure 6**. Dual-band MWP filters. Measured RF amplitude response of three dual passband filters and one dual stopband filter, including two dual passband filters with equal passband bandwidth, one with (a) 0.5 GHz and the other with (b) 1.5 GHz, (c) one dual passband filter with passband bandwidth of 4 GHz and 1.5 GHz, and (d) a dual stopband filter with two stopbands each having a bandwidth of 2 GHz.

Dual-band microwave filters have also attracted great interest because of the growing demand for multifunctional microwave systems supporting various modern applications [38-40]. Four distinct dual-band MWP filters were demonstrated here (**Figure 6**), further verifying the reconfigurability of the MWP filters. An 8.2-km SMF was employed, resulting in an operational bandwidth of ~8.84 GHz. Tap coefficients were designed to achieve three dual passband filters and one dual stopband filter, including two dual passband filters with equal passband bandwidth − one with 0.5 GHz (**Figure 6(a)**) and the other with 1.5 GHz (**Figure 6(b)**), one dual passband filter with passband bandwidth of 4 GHz and 1.5 GHz (**Figure 6(c)**), and a dual stopband filter with two stopbands each having a bandwidth of 2 GHz (**Figure 6(d)**).

**Figure 6** shows the RF amplitude response of the four filters. The roll-off rates of these filters were measured to be ~[10.7, 10.3, 20.0, 21.9] dB/GHz, ~[19.1, 20.2, 16.9, 9.3] dB/GHz, ~[12.8, 13.7, 10.1, 9.7] dB/GHz, and ~[9.9, 9.3, 7.2, 5.3] dB/GHz, respectively. The results show that dual-band MWP filters with various filter shapes can be achieved, verifying the high reconfigurability and consistent sharp roll-off performance of the proposed MWP filter.

To comprehensively examine the performance of MWP filters, RF metrics including spurious-free dynamic range (SFDR), RF link gain, and noise figure are adopted to characterize the RF performance of the system. To characterize the linearization of the proposed filter, we assessed the SFDR of a low-pass filter by using the two-tone method (**Figure 7(a)**) [41]. The RF response of the low-pass filter is shown in **Figure 7(b)** with a cut-off frequency of ~5 GHz. A two-tone test [14] at the frequency of 1 GHz was performed, where two CW RF signals at 1 GHz and 1.01 GHz were generated by an arbitrary waveform generator (AWG, Keysight) and then simultaneously fed to the IM with the same power. Through varying the RF input power of the two-tone signal from $-5.66$ dBm to $-0.43$ dBm and monitoring the output signals from the BPD on the electrical spectrum analyzer (ESA, Keysight), the amplitude of the third-order intermodulation distortion (IMD3) at 0.99 GHz and 1.02 GHz was recorded (**Figure 7(c)**), indicating a SFDR of ~91 dB·Hz$^{2/3}$ as shown in **Figure 7(d)**.

RF link gain of filter passbands reflects the overall signal transfer efficiency from the input to the output. The RF link gain at frequency $f$ can be described as [2]

$$G(f) = 10\log_{10}(P_{out} / P_{in}) \tag{6}$$

where $P_{out}$ and $P_{in}$ are the RF output and input power, respectively. The RF link gain of the proposed low-pass MWP filter (**Figure 7(b)**) was $\sim-14.8$ dB measured at 1 GHz, where the main sources of degradation are inefficient optical to electrical and electrical to optical conversion [42] as well as optical losses [2]. The optical losses can be compensated for by employing low-noise components and optical amplifiers. Noise figure quantifies the degradation of signal-to-noise ratio induced by system-inherent noise, which can be expressed by [2]

$$NF = 10\log_{10}(SNR_{in} / SNR_{out}) \tag{7}$$

where $SNR_{in}$ and $SNR_{out}$ are the signal-to-noise ratio (SNR) of the input and output signals, respectively, with an assumption that only the thermal noise exists at the input. The noise figure of the low-pass filter in **Figure 7(b)** was ~19.7 dB, depending on the noise of the soliton crystal microcomb, optical and RF amplifiers, and BPD.

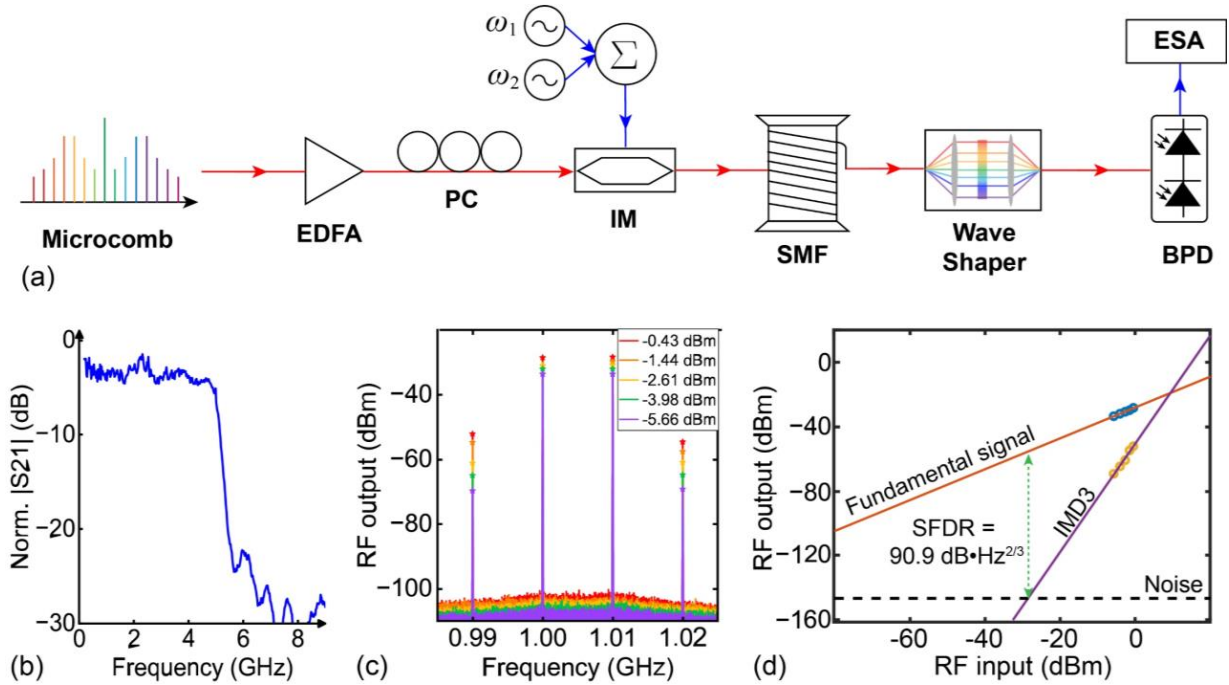

**Figure 7**. Linearization of the MWP filters. (a) Schematic of the experimental setup to demonstrate the linearization of the MWP filters. ESA: electrical spectrum analyzer. $\omega_1$ and $\omega_2$: angular frequencies of the two-tone input RF signal. (b) RF amplitude response of the proposed low-pass filter, with a cut-off frequency of ~5 GHz. (c) Measured two-tone RF output spectra at different input RF power. (d) Measured spurious-free dynamic range (SFDR) of the low-pass MWP filter. IMD3: third-order intermodulation distortion.

**Table 1** gives a summary of state-of-the-art MWP filters. For IIR MWP filters achieved by direct response mapping (DRM), although they feature simple structures and a high degree of integration, their roll-off performance largely depends on the sharpness of the employed optical filters. To address this limitation, cascaded MRRs have been introduced to achieve sharper optical filtering responses and thus enhance the roll-off performance of MWP filters after the conversion of phase modulation to intensity modulation (PM-IM) [7, 9]. FBGs have also been exploited after optical filters to attenuate phase modulated sidebands [3]. Although this led to enhanced shaped factors, the improvement sacrificed the reconfigurability of MWP filters due to the fixed spectral characteristics of FBGs. IIR MWP filters based on SBS in on-chip waveguides [43] or optical fibre [44] can achieve sharp transitions between the passband and stopband because of the narrow linewidth of the Brillouin gain spectrum, typically around tens of MHz [45]. However, this narrow linewidth constrains the tunability of the filter bandwidth [46].

In this work, we achieve FIR MWP filters based on the transversal filter system using a soliton crystal microcomb source, offering an ultrahigh roll-off performance and highly reconfigurable solution. By applying different tap coefficients and time delays, the system can synthesize arbitrary RF filter responses. We demonstrate four different types of single-band MWP filters as well as dual-band filters with various filter response, where both the bandwidth and centre frequency of the filters are tunable. In contrast to the approach adopted in our previous work [26, 34, 35], where equal tap coefficients were employed to realize low-pass filters first followed by additional modifications to tailor the filter response, we directly calculate the tap coefficients by applying IFFT to the desired RF amplitude response. This method facilitates improved roll-off performance, achieving a shape factor of ~1.15, by ensuring the system response more accurately aligns with the ideal response.

**Table 1.** Comparison of state-of-the-art MWP filters. SFDR: spurious free dynamic range. ROR: roll-off rate. DRM: direct response mapping. MRR: microring resonator. RAMZI: ring-assisted Mach-Zehnder Interferometer. DFBR: distributed feedback resonator. TOBF: tunable optical band-pass filter. FBG: fibre Bragg grating. EPS-FBG: equivalent phase-shifted FBG. SBS: simulated Brillouin scattering. PM-IM: phase modulation to intensity modulation. TFS: transversal filter structure. A: low-pass filter. B: band-pass filter. C: high-pass filter. D: band-stop filter. E: dual passband filter. F: dual stopband filter.

| Mechanism | Structure | Number of functions | Type of functions | Tuning range (GHz) | SFDR (dB·Hz$^{2/3}$) | ROR (dB/GHz) | Shape factor |
|---|---|---|---|---|---|---|---|
| DRM[7] | Two cascaded MRRs | 1 | B | $4.0 - 21.5$ | – | – | 1.23 |
| DRM[9] | Four cascaded MRRs | 1 | B | $4.0 - 36.0$ | – | 10.23 | – |
| DRM[47] | RAMZI | 1 | A | $0 - 6.0$ | 81.4 | – | – |
| DRM[6] | DFBR waveguide Bragg grating | 1 | B | $10.0 - 67.5$ | 95.8 | 7.0 | – |
| DRM[3] | TOBF + FBGs | 1 | B | $6.3 - 22.4$ | – | – | 1.88 |
| DRM[40] | EPS-FBG | 1 | E | $0.8 - 7.4$ | 89.8 | – | 3.3 |
| SBS[43] | Waveguide | 1 | D | $5.0 - 20.0$ | 92.2 | – | – |
| SBS[44] | Optical fibre | 2 | B, E | $1.65 - 2.15$ | – | – | 1.35 |
| TFS[26] | Microcomb | 4 | A, B, C, D | $2.0 - 6.0$ | – | – | – |
| TFS[34] | Microcomb | 1 | B | $3.28 - 19.4$ | – | 5.2 | – |
| This work | Microcomb | 6 | A, B, C, D, E, F | $0 - 25.0$ | 90.9 | 32.6 | 1.15 |

The roll-off performance is theoretically programmable but practically limited by the number of taps and the experimentally induced processing errors, including noise of the soliton crystal microcomb, chirp and uneven response of the IM, high-order dispersion in SMF, shaping errors of the Waveshaper, and noise of the BPD [35]. By reducing the linewidths and noise of lasers for pumping microcombs, using dispersive devices without high-order dispersion such as Bragg gratings, and employing advanced feedback control methods, there is still room for future improvement of the roll-off performance. The present low-pass MWP filter is with a SFDR of ~91 dB·Hz$^{2/3}$, which is comparable to those reported in other works. To increase the SFDR, an SBS assisted filter can be incorporated for carrier suppression [40].

These results confirm the effectiveness of optical microcombs in forming the basis for transversal filter microwave spectral filters [48 – 128] potentially involving advanced circuit designs [129 – 136] including graphene oxide and other 2D material based devices, [137 – 174] with applications to quantum optics. [175 – 190]

## 4. Conclusions

In this paper, we proposed ultrahigh-performance adaptive MWP filters based on a soliton crystal microcomb source. We demonstrated four single-band filters with distinct filter response, achieving extremely sharp roll-off rates up to ~32.6 GHz and a minimum 20-dB shape factor of ~1.15. To describe the tunability of the filters, band-pass filters with tunable centre frequencies ranging from 5GHz – 15 GHz were realized. In addition, we implemented dual-band filters with different filter shapes by using the same configuration, validating their high reconfigurability while maintaining high roll-off performance. The ultrahigh-performance adaptive filters presented in this paper offer a promising solution for the applications demanding high spectral selectivity and reconfigurability.

## Conflict of interests

The authors declare no conflict of interests.